\documentclass[usenatbib,usegraphicx,useAMS]{mn2e}
\usepackage{amssymb,url,times}

\def\be{\begin{equation}}
\def\ee{\end{equation}}

\title[Resolving fragmentation in SPH simulations]{Resolution requirements for Smoothed Particle Hydrodynamics simulations of self-gravitating accretion discs}

\author[Lodato \& Clarke]{Giuseppe Lodato$^1$\footnote{giuseppe.lodato@unimi.it} and C. J. Clarke$^2$\\
$^1$Dipartimento di Fisica, Universit\`a Degli Studi di Milano, Via Celoria, 16, Milano, 20133, Italy\\
$^2$ Institute of Astronomy, Madingley Road, Cambridge, CB3 0HA\\
}
\date{Submitted: Revised:  Accepted:}
\pagerange{\pageref{firstpage}--\pageref{lastpage}} \pubyear{2010}

\begin{document}
\label{firstpage}
\bibliographystyle{mn2e}
\maketitle

\begin{abstract}
Stimulated by recent results by \citet{meru10a,meru10b}, we revisit the issue of resolution requirements for simulating self-gravitating accretion discs with Smoothed Particle Hydrodynamics (SPH). We show that the results by \citet{meru10a} are consistent with those of \citet{meru10b} if they are both interpreted as driven by resolution effects, therefore implying that the resolution criterion for cooling gaseous discs is a function of the imposed cooling rate. We discuss two possible numerical origins of such dependence, which are both consistent with the limited number of available data. Our results tentatively indicate that convergence for current simulations is being reached for a number of SPH particles approaching 10 millions (for a disc mass of order 10 per cent of the central object mass), which would set the critical cooling time for fragmentation at about $15\Omega^{-1}$, roughly a factor two larger than previously thought. More in general, we discuss the extent to which the large number of recent numerical results are reliable or not. We argue that those results that pertain to the dynamics associated with gravitational instabilities (such as the locality of angular momentum transport, and the relationship between density perturbation and induced stress) are robust, while those pertaining to the thermodynamics of the system (such as the determination of the critical cooling time for fragmentation) can be affected by poor resolution.
\end{abstract}


\section{Introduction}

Gravitational instabilities in accretion discs have been intensively studied in the last decade. Several analytical models of gravitationally unstable discs have been proposed over the years \citep{BL99,balbus99,clarke09,rafikov09,rice10,krumholz10}. However, a large number of important properties of self-gravitating discs have been determined through the use of numerical simulations (\citealt{gammie01,LR04,LR05,RLA05,mejia05,boley06,cai06,boss04,boss06,CLC09,CLC10}, just to mention a few). Such numerical studies have investigated several different aspects of self-gravitating disc dynamics. Some (e.g. \citealt{LR04,LR05,boley06,CLC09}) have considered the angular momentum transport in non-fragmenting discs, with the aim of determining whether the long term disc evolution induced by the instability can be treated within a local framework. This is particularly relevant for constructing local analytical models of self-gravitating discs \citep{clarke09,rafikov09}. In some other cases \citep{gammie01,RLA05,boss04,mayer04,mayer05b,boss06,mayer07,CLC10}, the focus was on the conditions required for a self-gravitating disc to fragment into bound objects. This issue is extremely important in the context of planet formation theories, as this mechanism forms the basis of the so-called gravitational instability model for planet formation. More in general, such mechanism might be responsible for the formation of stellar clusters in galactic centers, including our own \citep{nayak06,nayak07}. The mechanism leading to fragmentation is fairly well understood. In a cooling disc, thermal saturation of the instability occurs when the amplitude of the density perturbation is large enough that the associated shock heating provided by the instability balances the cooling rate \citep{CLC09}. If the required density perturbation exceeds a critical value, the disc fragments. Determining the critical value for the density perturbation is thus equivalent to determining a critical cooling time below which fragmentation occurs. Such critical cooling time has been variously determined over the years: \citet{gammie01} sets it to $3\Omega^{-1}$ (where $\Omega$ is the local angular velocity of the disc), based on his two dimensional shearing sheet simulations, \citet{RLA05} set it to $6\Omega^{-1}$ for a specific heat ratio of $\gamma=5/3$, based on their three dimensional global SPH simulations, and note a dependency of the critical cooling time on $\gamma$. Other studies have found other dependencies either on the specific form of the cooling rate \citep{johnson03,CLC10} or on the thermal history of the disc \citep{CHL07}. It is convenient to express the cooling time in units of the local dynamical time in the disc and introduce a ``cooling parameter'' $\beta=\Omega t_{\rm cool}$. 

Recently, some new results have cast doubts on the reliability of the determination of such critical cooling time, and in particular on those obtained by SPH simulations. In particular, we consider here two new results: first, \citet{meru10a} have run a number of simulations with different slopes of the disc surface density profile, different disc masses, and have considered the location at which the fragment appeared in their fragmenting runs. Their conclusion is that the critical value of $\beta$ at a given radius $R$ is a function of the ``local disc mass'',  $m(R)=\Sigma(R) R^2/M_{\star}$, where $\Sigma$ is the surface density and $M_{\star}$ is the mass of the central object. In particular, they find that the critical cooling parameter for fragmentation increases roughly as $m^{1/2}$. Secondly, \citet{meru10b} have noted that, for a given disc set up (in terms of total disc mass, surface density profile, etc.) the critical $\beta$ for fragmentation is an increasing function of the total number of SPH particles used, and --- more importantly --- they find no hint of convergence. 

In this paper, we reconsider the issue of resolution requirements in order to simulate fragmentation in a cooling gaseous discs. We introduce a new resolution requirement that relates the number of SPH particles used to the externally imposed cooling rate. We then show that the whole set of results by \citet{meru10a,meru10b} can be naturally explained as a consequence of failing to satisfy this condition. Since the \citet{meru10a,meru10b} simulations are among the highest resolution SPH simulations of fragmenting discs to date, this failure is shared also by all previous SPH simulations. We will then discuss which of the previous results are reliable and which should be considered with more care. 

\section{Resolution requirements for self-gravitating cooling discs}

Several papers in the past have investigated the resolution requirements for resolving gas fragmentation in numerical simulations. In particular, \citet{bate97} require that the mass of SPH particles within a smoothing sphere be less that the Jeans mass. A similar condition, applicable also to grid-based simulations \citep{truelove97}, is that the minimum resolvable length (i.e. the mesh size or the smoothing length $h$ for grid-based and SPH simulations, respectively) be less than the Jeans length, although note that failing to satisfy the Truelove condition in grid-based simulations generally induces spurious fragmentation, while in SPH poor resolution may well lead to suppression of fragmentation. \citet{nelson06} lists three different conditions: the first is that adaptive gravitational softening is used (this is actually usually done in most modern SPH simulations, including those discussed here), the second is a variant of the \citet{bate97} and \citet{truelove97} conditions, and the third is that the smoothing length $h$ be less than the disc thickness $H$. However, also this last condition is equivalent to the \citet{bate97} and \citet{truelove97} condition for a marginally stable disc, for which the Jeans length is actually equal to the disc thickness. All these conditions can therefore be cast in the simple requirement:
\be
\frac{h}{H}\lesssim 1.
\ee

The natural measure of resolution for an SPH simulation of an accretion disc is thus the quantity $h/H$. In order to discuss the outcome of gravitationally unstable discs we are interested in evaluating this quantity at the onset of the instability, that is in a condition where the disc is marginally stable. Let us then estimate the ratio $h/H$ for a marginally stable disc. We first recall than marginal stability for a Keplerian disc implies that
\be
Q=\frac{c_{\rm s}\Omega}{\pi G \Sigma}\approx 1,
\ee
where $c_{\rm s}$ is the sound speed. This is equivalent to
\be
\frac{H}{R} = \pi Q\frac{\Sigma R^2}{M_{\star}} = \pi Q m(R),
\label{eq:sr}
\ee
where $m(R)=\Sigma R^2/M_{\star}$. 

The SPH smoothing length is set by the condition
\be
\rho h^3 = \eta^3 m_{\rm p},
\label{eq:smooth}
\ee
where $\rho$ is the local SPH density, $\eta$ is a numerical parameter generally set to 1.2 and $m_{\rm p}=M_{\rm disc}/N$ is the SPH particle mass, where $M_{\rm disc}$ is the total disc mass, and $N$ is the total number of SPH particles. With this prescription, a smoothing sphere contains roughly 60 SPH ``neighbours''. We evaluate the smoothing length at the disc midplane. The midplane density $\rho_{\rm c}$ is related to the surface density by $\rho_{\rm c}=\Sigma/2H$, which, inserted in Eq. (\ref{eq:smooth}), and using Eq. (\ref{eq:sr}) and the definition of $m(R)$, gives:
\begin{eqnarray}
\nonumber \frac{h}{H} & = & \frac{\eta}{m(R)}\left(\frac{2q}{\pi^2 Q^2 N}\right)^{1/3}\\
 & \approx & \frac{5\times 10^{-3}}{m(R)}\left(\frac{q}{0.1}\right)^{1/3}\left(\frac{N}{2.5\times 10^5}\right)^{-1/3},
 \label{eq:hoverh1}
\end{eqnarray}
where $q=M_{\rm disc}/M_{\star}$. Actually, if the disc is very poorly resolved, that is if $h/H>1$, the SPH method underestimates the midplane density by a factor $H/h$, and we have $\rho_{\rm c}\approx\rho(h/H)$. This corresponds to an equivalent 2D version of equation (\ref{eq:smooth}):
\be
\Sigma h^2 \approx 2\eta^3 m_{\rm p}.
\label{eq:smooth2}
\ee
In this poorly resolved case, one readily gets that $h/H$ is given by
\begin{eqnarray}
\nonumber \frac{h}{H} & \approx & \left(\frac{\eta}{m(R)}\right)^{3/2}\left(\frac{2q}{\pi^2 Q^2 N}\right)^{1/2}\\
 & \approx & \frac{3.4\times 10^{-4}}{(m(R))^{3/2}}\left(\frac{q}{0.1}\right)^{1/2}\left(\frac{N}{2.5\times 10^5}\right)^{-1/2}.
 \label{eq:hoverh2D}
\end{eqnarray}

It is also useful to evaluate $h/H$ at the outer edge of the disc, $R_{\rm out}$. If the surface density profile is a power-law with respect to $R$ with index $p<2$, we have that $q=M_{\rm disc}/M_{\star}=2\pi m(R_{\rm out})/(2-p)$, from which we get (for $h/H<1$) 
\begin{eqnarray}
\nonumber \left.\frac{h}{H}\right|_{\rm out} & = & \frac{2\pi \eta}{2-p}\left(\frac{2}{q^2\pi^2Q^2N}\right)^{1/3}\\
 & \approx  & 0.3\left(\frac{q}{0.1}\right)^{-2/3}\left(\frac{N}{2.5\times 10^5}\right)^{-1/3},
 \label{eq:hoverh2}
\end{eqnarray}
where the last equality holds for $p=1$. The equivalent of (\ref{eq:hoverh2}) when $h/H>1$ is
\begin{eqnarray}
\nonumber \left.\frac{h}{H}\right|_{\rm out} & = & \left(\frac{2\pi \eta}{2-p}\right)^{3/2}\left(\frac{2}{q^2\pi^2Q^2N}\right)^{1/2}\\
 & \approx  & 0.16\left(\frac{q}{0.1}\right)^{-1}\left(\frac{N}{2.5\times 10^5}\right)^{-1/2}.
 \label{eq:hoverh2_2D}
\end{eqnarray}
Note however that equation ({\ref{eq:hoverh2_2D}) is of little practical use, since in order to have $h/H>1$ at the disc outer edge, the number of SPH particles needs to be very small, $N\lesssim 6000$. 

We can now reconsider the results of \citet{meru10a,meru10b} in the light of Equations (\ref{eq:hoverh1}) and (\ref{eq:hoverh2}) (or (\ref{eq:hoverh2D}) and (\ref{eq:hoverh2_2D})). In fact, on the one hand \citet{meru10a} find that, for a given number of particles, there is a relation between the imposed value of $\beta$ and the quantity $m(R)$, evaluated at the location of the first fragment appearing in their simulations. On the other hand, \citet{meru10b} demonstrate that, for a given disc setup, the critical value of $\beta$ for fragmentation is an increasing function of $N$. Both results can be thus explained if (a) fragmentation is suppressed for low resolution and (b) the resolution requirement, as measured by the parameter $h/H$, is a function of the imposed cooling time. In this picture, \cite{meru10a} thus determine the location at which, for a given $N$, the resolution requirement is satisfied, while \citet{meru10b} determine the minimum $N$ at which, for a given setup, the requirement is satisfied. 
 
 In fact, it would be puzzling to interpret the results of \citet{meru10a} otherwise. By construction, such simulations are self-similar and, as long as global effects do not play a role (which is the case when $m(R)\lesssim 0.1$), there should be no preferred location in the disc. The only possible mechanism that breaks the self-similarity is in fact resolution. 

Stated otherwise, we can say that in order to detect fragmentation in a cooling disc, the resolution requirement must depend on $\beta$, in such a way that discs with longer cooling times require a better resolution (that is, a smaller $h/H$). In this case, fragmentation would then first occur at the minimum radius for which such resolution requirement is satisfied, since the dynamical and cooling
times are shorter than at larger radius. We can express such a condition in the form
\begin{equation}
\beta<\beta_{\rm res} \equiv \beta_{\rm res}(h/H)
\label{eq:fit}
\end{equation}

What is the actual functional relation between $\beta_{\rm res}$ and $h/H$? 
We explore this by plotting in Fig. \ref{fig:rfrag} the values of $\beta$ in the
fragmenting simulations of \citet{meru10a} versus $H/h$ at the
point of fragmentation, and assuming that the correlation in these two quantities 
is purely driven by resolution, that is that $\beta=\beta_{\rm res}$.  
The quantity $H/h$ is computed using equation (\ref{eq:hoverh1}) or (\ref{eq:hoverh2D}), 
and using the data provided by \citet{meru10a} in their table 2. 
We also indicate with a leftward pointing error bar the cases where
our estimate of $m(R)$ (which is based on the initial disc surface density
profile) is an over-estimate: in these cases fragmentation occurs
close to the disc inner edge and the surface density has been significantly
depleted by boundary effects by the stage of fragmentation.

There are evidently a variety of functional forms that will pass through
the data - we are hampered both by the fact that there is considerable
scatter and by the fact that we have no simulation points with $H/h > 3$.
One possibility is that there is a linear relationship between $\beta$ and
$H/h$ (see Section 3 for a discussion of this form in terms of the
role of artificial viscosity). We have fitted the data  with a least squares 
fit (shown with the solid line in Fig. \ref{fig:rfrag}) and have obtained:

\begin{equation}
\beta_{\rm res} \approx 2\left(\frac{h}{H}\right)^{-1}.
\label{eq:blinfit}
\end{equation}

\begin{figure}
\begin{center}
\includegraphics[width=\columnwidth]{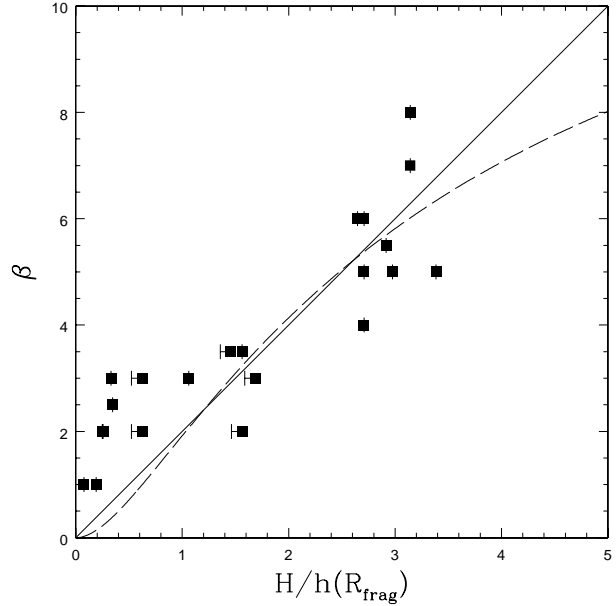}
\caption{Correlation between linear resolution at the location of fragmentation, as measured by $h/H$, and the imposed cooling time, as measured by the parameter $\beta$. Data points are taken from table 2 of \citet{meru10a}. A leftward pointing error bar indicates cases where fragmentation occurred near the disc inner boundary. The solid line shows a simple least squares fit to the data, assuming a linear relation (Eq. (\ref{eq:blinfit})). The best fit model in this case is $\beta=\beta_{\rm res}=2(H/h)$. The dashed line shows a least squares fit to the data assuming Eq. (\ref{eq:bquadfit}). The best fit parameters are $\beta_0=14.7$ and $a=1.77$.}
\label{fig:rfrag}
\end{center}
\end{figure}

It is worth stressing that if the trend $\beta=\beta_{\rm res}$, given by equation (\ref{eq:blinfit}),
extended  to indefinitely high $H/h$  then it would mean that there would be no {\it physical} limit 
to the cooling rate required for fragmentation. Instead 
it would imply that - however slow the cooling rate in the
disc - it would always fragment if modeled at high enough resolution. In contrast,
if there is indeed a physical limit to the cooling rate required for fragmentation 
then one would expect that at high $H/h$ the data would start deviating from the relation $\beta=\beta_{\rm res}$ and will saturate at a finite value of $\beta$. 
 
Alternatively, the dashed line shows a fit to the data of the form
\be
\beta_{\rm res} = \frac{\beta_0}{(1 + {ah}/{H})^2}
\label {eq:bquadfit}
\ee
(see Section 3 for a discussion of the numerical effects that could give
rise to such a relation). The important difference between this
functional form and that given by equation (\ref{eq:blinfit}) is that it
implies that $\beta_{\rm res}$ converges to a finite value ($\beta_0$)
at high $H/h$. The best fit values of the parameters are $\beta_0 = 14.7$ and $a=1.77$. 
We have excluded from the fit the data points that correspond to completely unresolved situations, that is $h/H>1$. 

 Evidently, either of the lines
shown in Figure 1 provide an adequate fit to the data over the
limited dynamic range of $H/h$ available.{\footnote{We 
also note that the data is adequately fitted by the
suggested parameterisation of Meru \& Bate (2010a), in which $\beta_{\rm res}$
(in scaling with $m(R)^{0.5}$) effectively scales with $(H/h)^{0.5}$. It is
also worth noting  the slightly different dependence of our  coordinate
$H/h$ on stellar mass compared with that of Meru \& Bate: we note that all
disc quantities depend on the central object mass only via the
Keplerian angular velocity, $\Omega$ and that consequently $\beta_{\rm res}$
must depend on $m(R)$ and $q$ in the combination $m(R)/q^{1/3}$.}}

We now turn our attention to the results of \citet{meru10b}, who had found an increase in the critical $\beta$ for fragmention at any radius with $N$. These authors consider the case $p=1$ and $q=0.1$. In the case where we are simply interested in fragmentation and not in the exact location within the disc where this happens, we only need to make sure that the resolution criterion is satisfied at the disc outer edge, that is
\begin{equation}
\beta<\beta_{\rm res}(h/H)_{\rm out} \approx 6.6\left(\frac{q}{0.1}\right)^{2/3}\left(\frac{N}{2.5\times 10^5}\right)^{1/3},
\label{eq:fitout}
\end{equation}
where the last approximate equality holds in the case that we adopt the linear fit (equation \ref{eq:blinfit}), and where we have used  equations (\ref{eq:hoverh2}) and (\ref{eq:fit}).

The data points of \citet{meru10b} are plotted in Fig. \ref{fig:reso}, along with our prediction for $\beta_{\rm res}$ 
based on Eq. (\ref{eq:fitout}) (solid line) and also the corresponding relationship based on equation (\ref{eq:bquadfit}) (dashed line).
We obtain the threshold for fragmentation $\beta_{\rm frag}$ from the data of \citet{meru10b} by averaging the highest value of their fragmenting runs with the lowest non fragmenting run. For this purpose, we consider as non fragmenting those runs that were classified as `borderline' by \citet{meru10b}.   

  We immediately see that the results of \citet{meru10a} and \citet{meru10b}
are mutually consistent if both sets of results are explained in terms of
resolution effects (that is, if $\beta_{\rm frag}=\beta_{\rm res}$), regardless of the exact parameterisation that is used
to describe how the cooling rate for fragmentation depends on $H/h$. 
We also note that, for the highest number of particles adopted by \citet{meru10b}, the results appear to deviate from the $N^{1/3}$ (solid) line implied by the
linear fit (equation \ref{eq:blinfit}).   In these cases, we thus have non-fragmenting simulations that are well resolved according to this  criterion. 

The simulation results in Figure 2 are also broadly consistent with the
dashed line (constructed using equation (\ref{eq:bquadfit}))  which is a
parameterisation that implies a converged value of the required cooling
rate for simulations that are sufficiently well resolved. 

Figure 2, and especially the deviation from the $N^{1/3}$ dependence at high $N$, therefore contains tentative evidence  that convergence may be  being reached at the largest number of particles, and that the 'true' value of the threshold cooling time for fragmentation would therefore lie between $\beta\approx 10-15$, that is roughly a factor of two larger than previously thought. 

\begin{figure}
\begin{center}    
\includegraphics[width=\columnwidth]{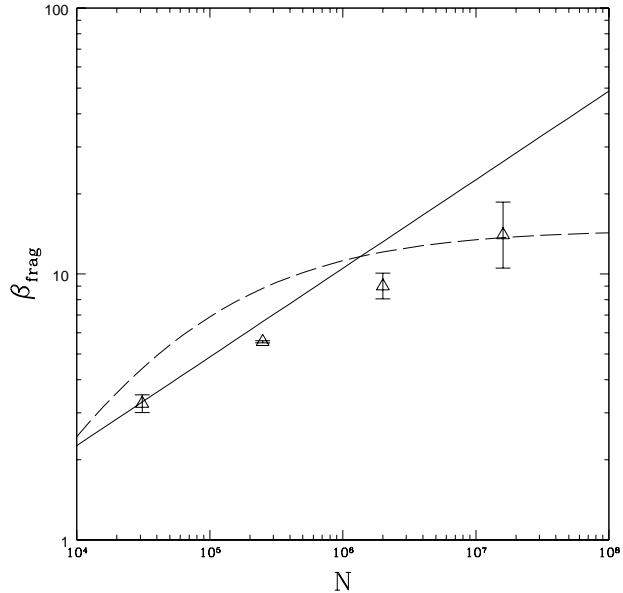}
\caption{Correlation between the critical cooling time for fragmentation and the number of SPH particles used, for the simulations described in \citet{meru10b}. The threshold is obtained averaging the highest $\beta$ for fragmenting simulations and the lowest $\beta$ for non-fragmenting ones. The solid and dashed lines show equations (\ref{eq:blinfit}) and (\ref{eq:bquadfit}), respectively, evaluated at the disc outer edge, and where the free parameters are set by the fit shown in Fig. \ref{fig:rfrag}.}
\label{fig:reso}
\end{center}
\end{figure}

\section{Physical origin of the resolution condition}

In the previous section, we have explained the available simulation data
in terms of two alternative parameterisations of how the cooling
rate required for fragmentation may depend on the ratio of the
smoothing length to the disc scale height.
We now turn our attention to possible physical explanations of such  requirements. 

The first possibility is related to artificial viscosity. Indeed, the artificial viscosity term in SPH is linearly proportional to $h/H$ \citep{monaghan92}, and its effect might in principle alter the thermal balance of the disc.

As discussed above, for cooling discs the saturation of the gravitational instability is due to the balance of internal heating due to shocks induced by the instability itself and the externally imposed cooling. Clearly, if any other form of external heating is present, the saturation amplitude of the instability will be decreased and fragmentation might therefore be inhibited even for cooling parameters that would lead to fragmentation in the absence of external heating. For actual astrophysically relevant discs, such external heating sources might well be an important physical ingredient to be taken into account. On the other hand, in numerical simulations, where one wishes to determine the critical cooling rate, one has be sure that such external heating sources, and in particular numerical heating sources, are negligible. 

It is well known \citep{pringle81} that the condition of thermal equilibrium for a viscous disc can be easily expressed in terms of a relation between the viscosity parameter $\alpha$ \citep{shakura73} and the cooling parameter $\beta$:
\be
\alpha= \frac{4}{9\gamma(\gamma-1)}\frac{1}{\beta}.
\label{eq:alpha}
\ee

For SPH simulations, the most significant source of numerical heating is provided by artificial viscosity. In particular, it can be shown \citep{lubow96,murray96,LP10} that the linear term in the standard artificial viscosity implementation of SPH leads to an equivalent Shakura-Sunyaev parameter $\alpha_{\rm art}$ given by
\be
\alpha_{\rm art} = \frac{1}{10}\alpha_{\rm SPH}\frac{h}{H},
\ee
where $\alpha_{\rm SPH}$ is an SPH input parameter (usually set to 1, but reduced to 0.1 in the simulations discussed here). Actually, the above relation holds only in the continuum limit and if artificial viscosity is applied both in regions of converging and diverging flow. Usually, a switch enabling artificial viscosity only for convergent flows is used, and sometimes additional switches \citep{balsara95} are used, so that the above relation is actually an overestimate of the resulting viscosity. 

When $\alpha_{\rm art}$ becomes comparable to the value implied by Eq. (\ref{eq:alpha}), artificial heating offsets completely the imposed cooling, and no gravitational perturbations will be able to grow. This occurs when
\be
\beta\approx\frac{40}{9\gamma(\gamma-1)}\frac{1}{\alpha_{\rm SPH}}\left(\frac{h}{H}\right)^{-1}= 40\left(\frac{h}{H}\right)^{-1},
\label{eq:condition}
\ee
where the last equality refers to the case of $\gamma = 5/3$ and $\alpha_{\rm SPH}=0.1$. The resolution criterion proposed in the previous section (equations \ref{eq:fit} and  \ref{eq:blinfit}) thus corresponds to the case where artificial viscosity provides 5 per cent of the heating required for thermal equilibrium. 

  We draw attention to two aspects of this result. Firstly, it  would imply - 
rather surprisingly -   that artificial viscosity effects can prevent
fragmentation even when they contribute a tiny fraction ($ 5 \%$) of the
total thermal energy balance.  Secondly, it should be stressed that 
this condition is a necessary (but not necessarily a sufficient) criterion
for fragmentation. As noted in Section 2, there is tentative evidence from
Figure 2 that the points at the highest $N$ fall below the solid line -
in other words we have simulations that are not fragmenting even though
they are in the regime where the artificial viscosity contribution
is less than $5 \%$. We interpret this result as evidence that there is indeed 
a physical mechanism preventing fragmentation for small cooling rates.

In addition to the thermal effects described above, artificial viscosity might also dynamically stabilize the disc. The specific effect of viscosity in this regard depends on how viscosity scales with surface density and might also lead to a secular instability \citep{schmit95}. While this is beyond the scope of the present paper, such effects should be further investigated. 

  Since we find it rather surprising that artifical viscosity should
suppress fragmentation when it contributes such a minor component
to the disc's thermal balance, we also explore the hypothesis that
the effect of under resolution is one of artificially smoothing
density enhancements over  the SPH smoothing length $h$. In poorly
resolved simulations, where $h$ is not much smaller than the length scale 
of density peaks ($\lambda$), then this will suppress the peak amplitude 
of the resulting density fluctuations  by a factor of $1 + h/\lambda$ (assuming
that gravitational instabilities generate predominantly linear structures).  
According to \citet{CLC09}, the gravitational heating rate associated with 
modes of r.m.s. fractional amplitude $\Delta \Sigma/\Sigma$ is proportional 
to $(\Delta \Sigma/\Sigma)^2$ and therefore in thermal equilibrium this quantity 
scales with the cooling rate ($\propto \beta^{-1}$). If we  assume that in a perfectly
resolved calculation the peak amplitude would scale with the r.m.s. amplitude 
(but be degraded by a factor $1 + h/\lambda$ in the case of finite $h$) and if we 
furthermore associate fragmentation
with the peak fluctuations achieving a critical value of $\Delta \Sigma/\Sigma$
(of order unity) then it follows that the cooling rate required for fragmentation is 
increased in the case of simulations of finite $h$. Specifically we have
\begin{equation}
\beta_{\rm{res}} = \beta_0 (1 +  h/\lambda)^{-2} 
\label{eq:bquadfit2}
\end{equation}
where $\beta_0$ is the value required for fragmentation in the case of a well 
resolved simulation (i.e. small $h$). In estimating $\lambda$ we  note that the 
Jeans length in a disc with Toomre $Q$ parameter close to unity is just the vertical
scale height $H$; it therefore seems logical that $\lambda$ should scale
with $H$ and this motivates the form of equation (\ref{eq:bquadfit}).   

We note that both our resolution criteria do not abandon the notion that (in a well
resolved simulation) there is a critical cooling rate associated with
fragmentation but take into account that the fact that more vigorous cooling
(lower $\beta$) is required in the case of poorly resolved simulations.

\section{Discussion and conclusions}

In this paper we have shown that the results of \cite{meru10a} (concerning
the location at which fragmentation occurs in SPH simulations of  
cooling self-gravitating discs) are quantitatively consistent (if
interpreted as being driven by  resolution effects) with the results of
\cite{meru10b} in which it is shown that the critical cooling rate for
fragmentation depends on the number of particles $N$. Indeed the
results of each paper imply the results of the other and both
imply that the measured cooling rate for fragmentation is a function of $h/H$
(where $h$ is the SPH smoothing length and $H$ is the disc scale height).

 It is difficult to assign a precise functional form for the way that
the critical cooling rate depends on resolution, given the scatter in the
simulation data and the relatively small range in $H/h$ for which simulations
are currently available. We have explored two possibilities for effects
that may explain the simulation results. In one case we explore the possibility
that a necessary criterion for fragmentation to be  detected is that the
artificial viscosity contributes less than a certain fraction of the
thermal energy input (and associated angular momentum transport) in the
disc. We find that the results at low resolution can be explained in these terms
but that the requirement on the contribution from artifical viscosity
is suprisingly stringent (i.e. fragmentation occurs only where this
contributes less than $5 \%$ of the energy input to the disc). Alternatively,
we consider the possibility that finite resolution just smooths out density
peaks so that fluctuations that would collapse if well resolved do not
achieve a critical amplitude at finite $h$. 
  
  Irrespective of how one  explains how resolution affects 
the cooling  requirements  for fragmentation, an important issue -
as stressed by \citet{meru10b} - is whether there is indeed a
convergence in the required cooling rate at high resolution. In other
words, is there a level of resolution above which one recovers the
result that has been assumed hitherto - i.e. that there is a {\it physical} 
condition on the  cooling rate required for fragmentation? We evidently
cannot answer this question with calculations that have not attained
convergence (see Figure 2 where $\beta_{\rm frag}$ rises mildly with $N$ 
even at the highest $N$ values studied). However, Figure 2 contains
hints of approaching convergence. One way to see this is to compare
the simulation data with the solid line which corresponds to
requirement of a $5 \%$ contribution from aritifical viscosity.
The fact that the right hand points lie below this solid line (where
$\beta_{\rm frag}$ scales with $N^{1/3}$) is evidence that this condition
may be necessary but not sufficient for fragmentation. Similarly, the
simulation data is consistent with the dashed line in which the
actual $\beta$ required for fragmentation converges to a value of $\beta_0\approx 14.7$
at high resolution.

  Clearly the question of ultimate convergence will only be settled
by further simulations at higher $N$.   If convergence is {\it not}
ultimately attained then it would involve a radical re-think of
all our understanding of gravitationally unstable discs since it would
imply that {\it any} self-gravitating disc (however slowly  cooling) 
should in reality fragment. 
This would be surprising because, in the case of very slow cooling,
the r.m.s. mode amplitude required to achieve thermal balance should
be very low and it would be unexpected (though possible in
principle) that such a disc nevertheless
exhibited locally non-linear density fluctuations giving rise to
collapsing fragments.

 The more conservative conclusion (if convergence is ultimately attained)
is that we have simply under-estimated the critical $\beta$ value
hitherto. This would involve some quantitative adjustment of
previous conclusions. For example,   
if the critical cooling time for fragmentation is roughly a factor $2$ higher than had been previously
thought then this affects the location at which giant planets can form through gravitational
instability \citep{meru10b}: in the regime of ice cooling appropriate to the outer parts of protostellar discs,
this change would bring in the minimum radius for fragmentation by a factor $\sim 2^{2/9}$ \citep{clarke09,CLC10}. Although this change is quite modest, it may be important to the discussion of whether recently imaged planets (e.g., 
around HR 8799, \citealt{marois08}, around $\beta$ Pic, \citealt{lagrange09} or around Fomahault, \citealt{kalas08}) 
could have been formed by gravitational instability. Note, however, that --- for a given critical cooling time --- the uncertainty in the determination of the fragmentation radius due to uncertainties in the relevant opacities, in the detailed vertical structure of the disc and due to the effects of magnetic fields are probably larger than the modest change discussed above. 

More in general, we might ask which ones of the properties of non fragmenting self-gravitating accretion discs are going to be significantly affected by poor resolution, in the sense expressed by Eq. (\ref{eq:fit}). Probably the most important property, apart from fragmentation, is related to the stress induced by the instability, and in particular its magnitude and the  locality 
of the associated dissipation (which are in turn related to the power spectrum of the modes excited by the instability). 

As mentioned in the Introduction, several studies \citep{LR04,LR05,CLC09} have considered the issue of locality of the transport induced by the instability. Here the question is whether the typical wavelength of the disturbances is a significant fraction of the radius $R$. The above studies (and in particular \citealt{CLC09}) have shown that the spectrum of the disturbances peaks at a wavelength $\lambda \approx H$ and becomes negligible for $\lambda\gtrsim 3H$. In this respect, thus, the relevant resolution requirement is the generally less restrictive condition $h/H\lesssim 1$ already discussed in \citet{nelson06}. Furthermore, not resolving the smallest scales would only prevent the disc from developing small scale structures, which would not contribute to global transport anyway. 

Let us now discuss whether the evaluation of the stress from such simulations is affected by Eq. (\ref{eq:fit}). Clearly, \emph{for a given disturbance}, regardless of the way it has been excited, a correct evaluation of the resulting stress only requires that the typical wavelengths of the disturbance are resolved. However, here the question is whether the disc response to an external boundary condition (in this case, to an externally imposed cooling rate) depends on resolution.
The condition of thermal equilibrium relates in a unique way the external cooling to the magnitude of the stress induced by the instability (eq. (\ref{eq:alpha})), under the hypothesis that the only heating source in the disc is provided by the instability itself and is dissipated locally. Here, we have shown that the new empirically determined resolution condition corresponds to the case where artificial viscosity only provides 5 per cent of the heating necessary for thermal balance. Therefore, unless the effects of artificial viscosity are much larger than expected, one would not expect a change in the stress above the 5 per cent level. In this respect, the effect of poor resolution is just to suppress the development of small scale disturbances, and thus reduce the peak value of the density perturbation, while keeping the overall average value of the stress essentially unchanged. 

Finally, it is worth noting that the discussion above is focussed on SPH simulations. Many previous results \citep{gammie01,mejia05,boley06}, including several determinations of the fragmentation boundary \citep{gammie01,mejia05,boss04,boss06}, have been obtained using grid-based codes. Artificial viscosity is also present to various degrees is grid-based codes, and one might thus conclude that an analogous resolution requirement should be formulated also in thus case. Whether also grid-based simulations do not converge in the determination of the critical time scale for fragmentation and the exact form of the resolution requirement in this case are questions that still need to be worked out. 

\section*{Acknowledgements}

We thank Phil Armitage, Matthew Bate, Farzana Meru and Ken Rice for interesting discussions. We also thank Charles Gammie, for an insightful referee's report.

\bibliography{lodato}

\label{lastpage}
\end{document}